\documentclass[eqsecnum,showpacs,amsmath,amssymb,superscriptaddress,nofootinbib,preprintnumbers]{revtex4-2}

\usepackage{graphicx}% Include figure files
\usepackage{dcolumn}% Align table columns on decimal point
\usepackage{bm}% bold math
\usepackage{ulem}
\usepackage{color}

\newcommand{\bea}{\begin{eqnarray}}
\newcommand{\eea}{\end{eqnarray}}

\newcommand{\hide}[1]{}

\begin{document}

\preprint{NITEP 278}
\preprint{NU-QG-18}

\title{
%Quantum-Improved Charged Black Holes with Consistent Thermodynamics//
Consistency in the Quantum-Improved Charged Black Holes
}

\author{Chiang-Mei Chen}
\email{cmchen@phy.ncu.edu.tw}
\affiliation{Department of Physics, National Central University, Zhongli, Taoyuan 320317, Taiwan}
\affiliation{Center for High Energy and High Field Physics (CHiP),
National Central University, Zhongli, Taoyuan 320317, Taiwan}
\affiliation{Department of Physics and Kobayashi-Maskawa Institute, Nagoya University, Nagoya 464-8602, Japan}
\affiliation{Department of Physics, Rikkyo University, Toshima, Tokyo 171-8501, Japan}

\author{Akihiro Ishibashi} \email{ishibashi.akihiro.r7@f.mail.nagoya-u.ac.jp}
\affiliation{Department of Physics and Kobayashi-Maskawa Institute, Nagoya University, Nagoya 464-8602, Japan}

\author{Nobuyoshi Ohta} \email{ohtan.gm@gmail.com}
\affiliation{Nambu Yoichiro Institute of Theoretical and Experimental Physics (NITEP),
Osaka Metropolitan University, Osaka 558-5585, Japan}
\affiliation{Department of Physics and Astronomy, Kwansei Gakuin University, Sanda, Hyogo 669-1330, Japan}

\date{\today}

\begin{abstract}
We investigate the consistency in the thermodynamics and the approaches at the equation and action levels for the quantum-improved charged black holes with scale-dependent couplings. 
For the quantum-improved Reissner–Nordstr\"om black holes, we find that the thermodynamic consistency allows both the Newton and electromagnetic couplings to have arbitrary dependence on the radial coordinate.
We point out a subtlety in the chemical potential with the scale-dependent electromagnetic coupling in the study of thermodynamics.
We also examine the compatibility of the Einstein equations at the equation and action levels with the Bianchi identity, identifying the need for an additional quantum energy-momentum tensor. We then find that the consistency between the approaches at the equation and action levels requires that the Newton coupling satisfy certain property. Finally, we extend the analysis to cosmological solutions, suggesting that quantum-induced modifications can drive the isotropization of the early universe.

\end{abstract}

\maketitle

%%%%%%%%%%%%%%%%%%%%%%%%%%%%%%%%%%%%%%%%%%%%%%%%%%%%%%%%%%%%%%%%%%%%%%
\section{Introduction}
\label{introduction}
%%%%%%%%%%%%%%%%%%%%%%%%%%%%%%%%%%%%%%%%%%%%%%%%%%%%%%%%%%%%%%%%%%%%%%

Einstein's general relativity, which describes gravitational interactions, is widely regarded as a classical theory. Its most important physical predictions to cosmological and black hole solutions inevitably contain spacetime singularities. It is generally believed that such singularities will be resolved once quantum effects are properly taken into account. Various approaches to quantum gravity have been proposed, such as string theory~\cite{Polchinski:1998} and loop quantum gravity~\cite{Rovelli:1997yv}. However, a fully satisfactory and universally accepted theory remains elusive. In this situation, many efforts have been made to incorporate quantum effects in an effective manner, for instance through running couplings governed by renormalization group equations with nontrivial ultraviolet (UV) fixed points~\cite{Reuter:1996cp, Souma:1999at, Niedermaier:2006wt, ASrev1, ASrev2, ASrev3}. Here we discuss the problem using such running couplings in more general context and call these couplings scale-dependent ones.

Various aspects of black holes have been discussed with such scale-dependent couplings~\cite{BR:2000, BR:2006, FLR12, Reuter:2010xb, KS141, PS:2018, Ishibashi:2021kmf}.
In particular, a wide range of regular black hole solutions have been proposed~\cite{Lan:2023cvz}, starting from generalizations of the Schwarzschild geometry and extending to rotating and/or charged cases; representative examples include Refs.~\cite{Ayon-Beato:1998hmi, Ansoldi:2006vg, Bambi:2013ufa, Balart:2014cga, Toshmatov:2014nya, Lemos:2016ulj, Gaete:2022ukm}. However, for solutions characterized by two or more physical parameters, the issue of thermodynamic consistency has largely been overlooked. In many works, it is implicitly assumed that black hole thermodynamics is automatically preserved, and explicit consistency checks are often absent.

In fact, potential inconsistencies in the quantum-improved first law have been noted in the literature. For example, in~\cite{Reuter:2010xb, Ruiz:2021qfp}, it was proposed to redefine the Hawking temperature in order to restore thermodynamic consistency. Nevertheless, such a redefinition is questionable, since the Hawking temperature also admits an independent physical interpretation via the Unruh effect and is directly proportional to the surface gravity.

The black hole thermodynamics is fundamentally a quantum phenomenon and should therefore be preserved in any quantum-improved black hole solutions. From this perspective, maintaining thermodynamic consistency is not merely a technical requirement but a fundamental criterion for physically viable regular black holes, especially in the rotating or charged cases. Unfortunately, this important aspect has been overlooked in most of the solutions proposed in the literature.

In Ref.~\cite{Chen:2022xjk}, we first investigated the issue of thermodynamic consistency for rotating black holes within the framework of asymptotically safe gravity. We found that consistency is preserved only when the scale-dependent Newton coupling is a function of the black hole area evaluated at the horizon. This result provides a strong and nontrivial constraint on viable black hole solutions. The scale-dependent Newton coupling can also be equivalently interpreted as an effective mass distribution with a constant (classical) Newton constant. Our findings therefore offer a useful criterion for testing the thermodynamic consistency of other proposed regular rotating black hole solutions. Furthermore, regular rotating black holes with consistent thermodynamics were constructed in Ref.~\cite{Chen:2023wdg}, and their implications for black hole shadows were explored in Ref.~\cite{Chen:2025jay}.

In the present work, we extend this analysis to the charged black holes within the asymptotically safe gravity at three distinct levels: the solution, field equation, and action levels\footnote{While many previous studies have been performed at the solution level as mentioned above, a more fundamental understanding requires studies at both the solution and action levels. In particular, to discuss the consistency with black hole thermodynamics, it is worth studying the action level quantum improvement, since a general formulation of black hole thermodynamics has been established via the Noether charge method starting from a diffeomorphism invariant Lagrangian~\cite{Wald:1993nt,Iyer:1994ys}.}. At the solution level, we begin by promoting both the Newton gravitational constant and the electromagnetic coupling to the scale-dependent quantities in the classical Reissner–Nordstr\"om (RN) solution. Unlike the rotating case, charged black holes involve two independent couplings, providing additional freedom to satisfy the consistency condition. We show that allowing both couplings to depend on the radial coordinate is sufficient to ensure thermodynamic consistency. However, an important subtlety arises: 
For RN black holes with scale-dependent couplings, as will be shown in Subsec.~\ref{sec_solution}, the charge $Q$ and the scale-dependent Maxwell coupling $e(r)$ enter the metric function and the chemical potential through the combinations $Q^2/e(r)^2$ and $Q/e(r)^2$, respectively. We demonstrate that this particular structure guarantees thermodynamic consistency. In contrast, if one instead introduces an arbitrary radial charge distribution through $Q$, as in Refs.~\cite{Ansoldi:2006vg, Lemos:2016ulj, Gaete:2022ukm}, such consistency is generally no longer preserved. On the other hand, for regular charged black holes constructed within nonlinear electrodynamics, for example in Refs.~\cite{Ayon-Beato:1998hmi, Balart:2014cga}, the issue of thermodynamic consistency likewise is highly nontrivial and must be carefully examined.

We then examine the situation in the equation and action levels, where we derive the quantum improved Einstein equations by consistently incorporating the scale-dependent couplings. In both formulations, it is necessary to introduce an appropriate quantum energy–momentum tensor to ensure compatibility of the Einstein equations with the Bianchi identity. It has been argued that if the consistency is required for {\it any} Newton coupling, then this uniquely leads to Brans-Dicke type theory at the action level~\cite{Reuter:2003ca, Reuter:2004nv}. However we find that this brings back the theory classical one without quantum effects in the Einstein frame. In contrast, we show that if we relax the requirement that the consistency be valid for some {\it given} Newton coupling, it is possible to have consistent quantum improvement both at the equation and action levels. As one of the possible solutions of the consistency, we consider the case that $1/G$ is a harmonic function.
Finally, we attempt to solve the resulting quantum-improved Einstein equations and analyze the associated quantum effects emerging from these solutions.

The paper is organized as follows. In Sec.~\ref{sec_classical}, we review the thermodynamic consistency of classical charged black hole solutions. Quantum-improved charged black holes are then discussed in Sec.~\ref{sec_quantum}, where quantum effects are incorporated at three different levels: the solution level in Subsec.~\ref{sec_solution}, the equation level in Subsec.~\ref{sec_equation}, and the action level in Subsec.~\ref{sec_action}. At both the equation and action levels, it is pointed out that a quantum energy-momentum tensor is necessary to ensure that the resulting field equations remain compatible with the Bianchi identity. We also discuss how the approach at the equation level may be compatible with that at the action level. The interior geometry of the quantum-improved charged black holes considered here corresponds to a Bianchi-I cosmology, which is discussed in Sec.~\ref{sec_cosmology}. Finally, the main results are summarized in Sec.~\ref{sec_conclusion}.

%%%%%%%%%%%%%%%%%%%%%%%%%%%%%%%%%%%%%%%%%%%%%%%%%%%%%%%%%%%%%%%%%%%%%%
\section{Classical Charged Black Holes}
\label{sec_classical} 
%%%%%%%%%%%%%%%%%%%%%%%%%%%%%%%%%%%%%%%%%%%%%%%%%%%%%%%%%%%%%%%%%%%%%%

Let us consider the action of the classical Einstein–Maxwell theory
\begin{equation}
\label{eq_action}
S = \int d^4x \sqrt{-g} \left( \frac{R}{16 \pi G_0} - \frac{1}{4 e_0^2} F_{\alpha\beta} F^{\alpha\beta} \right),
\end{equation}
where $G_0$ and $e_0$ denote the gravitational Newton constant and electromagnetic coupling constant, respectively. 
%In the classical framework, both couplings are taken to be constant, $G = G_0$ and $e = e_0$. 
Varying the action with respect to $g^{\mu\nu}$ and $A_\nu$ one can derive the Einstein and Maxwell equations 
\begin{equation}
\label{eq_fieldEQ}
R_{\mu\nu} - \frac12 R g_{\mu\nu} = \frac{8 \pi G_0}{e_0^2} T_{\mu\nu}, \qquad 
\nabla_\mu F^{\mu\nu} = 0,
\end{equation}
where the rhs of the Einstein equation is the energy-momentum tensor for the Maxwell field:
\begin{equation}
\label{eq_MaxwellT}
T_{\mu\nu} \equiv F_{\mu\alpha} F_\nu{}^\alpha - \frac14 F_{\alpha\beta} F^{\alpha\beta} g_{\mu\nu}.
\end{equation}

The classical charged black-hole solution describes a spherically symmetric spacetime together with a purely electric gauge field
\begin{equation}
\label{eq_metric}
ds^2 = - f(r) dt^2 + \frac{dr^2}{f(r)} + r^2 d\Omega_2^2, \qquad A = A_t(r) dt,
\end{equation}
where 
\begin{equation}
\label{eq_solution}
f(r) = 1 - \frac{2 G_0 M}{r} + \frac{G_0 \pi Q^2}{e_0^2 r^2},
\qquad
A_t(r) = \frac{Q}{r}.
\end{equation}

The horizon radius $r_+$ is determined as the largest root of $f(r_+) = 0$, namely 
\begin{equation}
r_+ = G_0 M + \sqrt{G_0^2 M^2 - G_0 \pi Q^2/e_0^2}.
\end{equation}
The Hawking temperature and chemical potential are then given by
\begin{equation}
T = \frac{f'(r_+)}{4 \pi} = \frac1{4 \pi} \left( \frac{2 G_0 M}{r_+^2} - \frac{2 G_0 \pi Q^2}{e_0^2 r_+^3} \right), \qquad
\Phi = \frac{\pi}{e_0^2} A_t(r_+) = \frac{\pi Q}{e_0^2 \, r_+}.
\end{equation}

The first law of black hole thermodynamics takes the form
\begin{equation}
dM = T dS + \Phi dQ,
\end{equation}
from which the entropy $S$ can be obtained via
\begin{equation}
dS = \frac1{T} dM - \frac{\Phi}{T} dQ.
\end{equation}
Consistency of this definition requires the integrability condition
\begin{equation} 
\label{eq_cons_MQ}
\partial_Q \left( \frac1{T} \right)_M = - \partial_M \left( \frac{\Phi}{T} \right)_Q.
\end{equation}
The classical charged black hole solution satisfies this consistency condition. However this is a nontrivial problem in the quantum-improved black holes.

%%%%%%%%%%%%%%%%%%%%%%%%%%%%%%%%%%%%%%%%%%%%%%%%%%%%%%%%%%%%%%%%%%%%%%
\section{Quantum-Improved Charged Black Holes} \label{sec_quantum} 
%%%%%%%%%%%%%%%%%%%%%%%%%%%%%%%%%%%%%%%%%%%%%%%%%%%%%%%%%%%%%%%%%%%%%%

Due to quantum effects, the couplings are expected to run with the energy scale. Upon identifying an appropriate cutoff, the scale-dependent couplings generally become functions of the radial coordinate and the physical parameters:
\begin{equation}
G_0 \to G = G(r; M, Q), \qquad e_0 \to e = e(r; M, Q).
\end{equation}
We are now going to examine the consistency of this ansatz and the thermodynamics.

\subsection{Quantum Improvement at Solution Level and Thermodynamics}
\label{sec_solution} 
%%%%%%%%%%%%%%%%%%%%%%%%%%%%%%%%%%%%%%%%%%%%%%%%%%%%%%%%%%%%%%%%%%%%%%

To incorporate quantum effects at the level of the solution, one may formally replace the constant couplings in the classical solution~\eqref{eq_solution} with the scale-dependent couplings. However, such a substitution does not automatically guarantee the consistency of black hole thermodynamics~\cite{Ruiz:2021qfp, Reuter:2010xb, Chen:2022xjk, Chen:2023wdg}. Instead, the consistency condition imposes nontrivial constraints on the allowed functional forms of the scale-dependent couplings.

For convenience, we introduce $\alpha = \pi/e^2$, which simplifies the analysis. The horizon radius then satisfies
\begin{equation}
\label{radius}
r_+^2 - 2 G M r_+ + G \alpha Q^2 = 0.
\end{equation}
Taking the total differential, we obtain
\begin{eqnarray}
&& \left( 2 r_+ - 2 G M - 2 M r_+ \partial_{r_+} G + Q^2 \alpha \partial_{r_+} G + Q^2 G \partial_{r_+} \alpha \right) dr_+ 
\nonumber\\
&-& \left( 2 G r_+ + 2 M r_+ \partial_M G - Q^2 \alpha \partial_M G - Q^2 G \partial_M \alpha \right) dM
\nonumber\\
&+& \left( 2 G \alpha Q - 2 M r_+ \partial_Q G + Q^2 \alpha \partial_Q G + Q^2 G \partial_Q \alpha \right) dQ = 0.
\end{eqnarray}
Using $dr_+ = \partial_M r_+ dM + \partial_Q r_+ d Q$ and requiring the coefficients of $dM$ and $dQ$ to vanish, we obtain
\begin{eqnarray}
\partial_M r_+ &=& \frac{2 G r_+ + 2 M r_+ \partial_M G - Q^2 \alpha \partial_M G - Q^2 G \partial_M \alpha}{2 r_+ - 2 G M - 2 M r_+ \partial_{r_+} G + Q^2 \alpha \partial_{r_+} G + Q^2 G \partial_{r_+} \alpha},
\label{eq_pMr} \\
\partial_Q r_+ &=& - \frac{2 G \alpha Q - 2 M r_+ \partial_Q G + Q^2 \alpha \partial_Q G + Q^2 G \partial_Q \alpha}{2 r_+ - 2 G M - 2 M r_+ \partial_{r_+} G + Q^2 \alpha \partial_{r_+} G + Q^2 G \partial_{r_+} \alpha}. 
\label{eq_pQr}
\end{eqnarray}
These relations allow us to reexpress the consistency condition~\eqref{eq_cons_MQ}, now implicitly including the $r_+$-dependence, as
\begin{equation}
\label{eq_cons_rMQ}
\partial_Q ( 1/T ) + ( \partial_Q r_+ ) \; \partial_{r_+} ( 1/T ) + \partial_M ( \Phi/T ) + (\partial_M r_+) \; \partial_{r_+} ( \Phi/T ) = 0.
\end{equation}

Due to the scale-dependent couplings, the Hawking temperature, determined by the surface gravity, is modified and receives additional contributions from the radial derivatives of $G$ and $\alpha$ as
\begin{eqnarray}
T &=& \frac1{4 \pi} \left( \frac{2 M (G - r_+ \partial_{r_+} G)}{r_+^2} - \frac{Q^2 (2 \alpha G - r_+ \alpha \partial_{r_+} G - r_+ G \partial_{r_+} \alpha)}{r_+^3} \right)
\nonumber\\
&=& \frac{2 r_+ - 2 M \partial_{r_+}( r_+ G) + Q^2 \partial_{r_+} (\alpha G) }{4 \pi r_+^2},
\end{eqnarray}
where we have used~\eqref{radius} in going to the second line.
Moreover, the scale-dependent electromagnetic coupling introduces a nontrivial relation between the chemical potential and the gauge potential
\begin{equation}
\Phi = \alpha(r_+) \frac{Q}{r_+}.
\end{equation}

One can readily verify that if both $G$ and $\alpha = \pi/e^2$ depend only on the radial coordinate $r_+$, the thermodynamic consistency condition is satisfied. In this case, Eqs.~\eqref{eq_pMr} and~\eqref{eq_pQr} reduce to
\begin{equation}
\partial_M r_+ = \frac{G}{2 \pi r_+ T}, \qquad \partial_Q r_+ = - \frac{G \alpha Q}{2 \pi r_+^2 T}.
\end{equation}
Substituting these expressions into the left-hand side of Eq.~\eqref{eq_cons_rMQ}, since $\Phi$ does not depend on $M$ explicitly, we obtain
\begin{eqnarray}
&& \partial_Q \left( \frac1{T} \right) - \frac{G \alpha Q}{2 \pi r_+^2 T} \; \partial_{r_+} \left( \frac1{T} \right) + \frac{\alpha Q}{r_+} \; \partial_M \left( \frac1{T} \right) + \frac{G Q}{2 \pi r_+ T} \; \partial_{r_+} \left( \frac{\alpha}{r_+ T} \right)
\nonumber\\
&=& - \frac1{T^2} \partial_Q T - \frac{\alpha Q}{r_+ T^2} \partial_M T + \frac{G Q}{2 \pi r_+ T^2} \; \partial_{r_+} \left( \frac{\alpha}{r_+} \right).
\end{eqnarray}
From the expression for the temperature, we further obtain
\begin{equation}
\partial_Q T = \frac{Q}{2 \pi r_+^2} \partial_{r_+} ( \alpha G), \qquad \partial_M T = - \frac1{2 \pi r_+^2} \partial_{r_+} (r_+ G).
\end{equation}
Using these relations, the consistency condition reduces to
\begin{equation}
\frac{Q}{2 \pi r_+^2 T^2} \left[ - \partial_{r_+} (\alpha G) + \frac{\alpha}{r_+} \, \partial_{r_+} (r_+ G) + r_+ G \; \partial_{r_+} \left( \frac{\alpha}{r_+} \right) \right] \equiv 0,
\end{equation} 
which is identically satisfied. Therefore, thermodynamic consistency is preserved for arbitrary functions $G(r_+)$ and $\alpha(r_+)$.

This result is consistent with expectations based on studies of quantum-improved rotating black holes~\cite{Chen:2022xjk}. In order to preserve the consistency of black hole thermodynamics, the scale-dependent couplings should depend on the horizon area. In a spherically symmetric spacetime, any function of the radial coordinate can equivalently be regarded as a function of the horizon area. It is then natural to consider this dependence extends away from the horizon.

From this perspective, any generalizations of charged black holes must preserve thermodynamic consistency by introducing an appropriately running electromagnetic coupling. This feature must be carefully taken into account when analyzing charged fields interacting with the electromagnetic background of the black hole. 

In particular, note that the charge $Q$ and the scale-dependent Maxwell coupling $e(r)$ enter the metric function and the chemical potential through the combinations $Q^2/e(r)^2$ and $Q/e(r)^2$, respectively. Since these terms involve different powers of $Q$, it is generally impossible to construct an equivalent description by simply introducing a prescribed radial charge distribution $Q(r)$~\cite{Ansoldi:2006vg, Lemos:2016ulj, Gaete:2022ukm}. Therefore, for regular charged black holes, establishing thermodynamic consistency remains a crucial and highly nontrivial issue that must be treated with particular care.

\subsection{Quantum Improvement at Equation Level} \label{sec_equation} %\label{QIeq}
%%%%%%%%%%%%%%%%%%%%%%%%%%%%%%%%%%%%%%%%%%%%%%%%%%%%%%%%%%%%%%%%%%%%%%

Another possibility is to incorporate quantum effects at the level of the field equations, i.e. by substituting the scale-dependent into the Einstein and Maxwell equations~\eqref{eq_fieldEQ} and then solve for the corresponding geometry. Note that this approach is not applicable to vacuum solutions such as the Schwarzschild or Kerr black holes, since the Newton coupling does not explicitly appear in the vacuum Einstein equations.

Because the Maxwell equations remain unchanged, the gauge potential retains the same form as in the classical solution:
\begin{equation}
A_t(r) = \frac{Q}{r}.
\label{qMaxwell}
\end{equation}
The energy-momentum tensor of the Maxwell field~\eqref{qMaxwell} is given by
\begin{equation}
T^\mu{}_\nu = \frac{Q^2}{8 r^4} \; \mathrm{diag}(-1, -1, 1, 1).
\end{equation}
However, the couplings entering the Einstein equations are no longer constant. As a result, consistency with the Bianchi identity and the conservation law requires the inclusion of an additional quantum energy-momentum tensor~\cite{Reuter:2003ca}:
\begin{equation}
\label{eq_QfieldEQ}
R_{\mu\nu} - \frac12 R g_{\mu\nu} = 8 \alpha G T_{\mu\nu} + \vartheta^\mathrm{(e)}_{\mu\nu}, 
\end{equation}
where $\vartheta^\mathrm{(e)}_{\mu\nu}$ satisfies 
\begin{equation}
\label{eq_vartheta_e}
\nabla^\mu \vartheta^\mathrm{(e)}_{\mu\nu} = - 8 T_{\mu\nu} \nabla^\mu ( \alpha G ).
\end{equation}
Since we are considering quantum theory, it is natural to expect that quantum effects produce some energy-momentum.

On the assumption that the quantum energy-momentum tensor is diagonal
\begin{equation}
\vartheta^\mathrm{(e)}{}^\mu{}_\nu = \mathrm{diag}\,\Bigl( -\rho_q(r), \, p_{q1}(r), \, p_{q2}(r), \, p_{q3}(r) \Bigr),
\end{equation}
Eq.~\eqref{eq_vartheta_e} yields only two nontrivial components. The $\nu = \theta$ component implies $p_{q3} = p_{q2}$, while the $\nu = r$ component reduces to
\begin{equation}
r^4 \left( \rho_q + p_{q1} \right) f'  - 2 \left[ Q^2 (\alpha G)' - r^3 ( r p_{q1}' + 2 p_{q1} - 2 p_{q2} ) \right] f = 0.
\label{detla}
\end{equation}
It is natural to assume that the quantum energy-momentum tensor does not depend explicitly on the detailed form of the lapse function $f(r)$. We find that if $p_{q1} = -\rho_q$, the lapse function drops out of Eq.~\eqref{detla}.
For simplicity, we also assume $p_{q2} = - w \, p_{q1} = w \, \rho_q$ with a parameter $w$. Our final form of the energy momentum-tensor is then
\bea
\vartheta^\mathrm{(e)}{}^\mu{}_\nu = \mathrm{diag}\,\Bigl( -\rho_q(r), \, -\rho_q(r), \, w \, \rho_q, \, w \, \rho_q \Bigr).
\label{perf_f}
\eea
The above equation~\eqref{detla} then reduces to
\begin{equation}
Q^2 (\alpha G)' + r^4 \rho_q' + 2 (1 + w) r^3 \rho_q = 0,
\label{eq02}
\end{equation}
whose formal solution is
\begin{equation}
\rho_q = r^{-2 w - 2} \left( c - \int \frac{Q^2}{r^{-2 w + 2}} (\alpha G)' dr \right),
\label{sol1}
\end{equation}
with an integration constant $c$.
The corresponding solution to the Einstein equations~\eqref{eq_QfieldEQ} leads to the lapse function
\begin{equation}
%\label{laps:equation-level}
f(r) = 1 - \frac{2 G_0 M}{r} - \frac1{r} \int \left( \frac{Q^2 \alpha G}{r^2} + r^2 \rho_q \right) dr.
\label{sol2}
\end{equation}
This expression already manifests a formal difference from the solution-level quantum improvement reported in Ref.~\cite{Ishibashi:2021kmf} (see also appendix). In particular, it possesses a central singularity at $r=0$, unless the third term cancels the effects of the second term. However, since our current setup differs from that of~\cite{Ishibashi:2021kmf}, one should be cautious about drawing any immediate physical conclusions from a simple comparison.

We now highlight two physically interesting cases:
\begin{enumerate}
\item [(i)] {\bf Maxwell-type ($w = 1$).}

In this case, the quantum energy--momentum tensor has the same structure as that of the Maxwell field,
\begin{equation}
\vartheta^\mathrm{(e)}{}^\mu{}_\nu = \rho_q(r) \;\mathrm{diag}(-1, -1, 1, 1),
\end{equation}
with 
\begin{equation}
\rho_q(r) = \frac{c - Q^2 \alpha G}{r^4}.
\end{equation}
The lapse function then simplifies to
\begin{equation}
f(r) = 1 - \frac{2 G_0 M}{r} + \frac{c}{r^2}.
\end{equation}
Thus, the quantum contribution exactly cancels that of the Maxwell field, and the solution reduces to the classical Reissner-Nordstr\"om form. However we should note that the origin of the last term is quantum effect, and this is different from the classical Reissner-Nordstr\"om solution.

\item[(ii)] {\bf Cosmological-constant-type ($w = -1$).}

For $w = -1$, the quantum energy--momentum tensor takes the form
\begin{equation}\label{theta:cc-type}
\vartheta^\mathrm{(e)}{}^\mu{}_\nu = \rho_q(r) \;\mathrm{diag}(-1, -1, -1, -1),
\end{equation}
with
\begin{equation}
\rho_q(r) = c - \frac{Q^2}{r^4} \alpha G - 4 Q^2 \int \frac{\alpha G}{r^5} dr.
\end{equation}
In this case, $\vartheta^\mathrm{(e)}_{\mu\nu}$ takes the form of a perfect fluid with equation of state $w = -1$, effectively behaving as a scale-dependent cosmological constant. The quantum energy density vanishes when either $Q = 0$ or $\alpha G$ is constant. The corresponding lapse function is then given by
\begin{equation}
\label{lapse_QI_eqlevel}
f(r) = 1 - \frac{2 G_0 M}{r} + \frac{4 Q^2}{r} \int r^2 \int^r \frac{\alpha(r') G(r')}{r'^5} dr' dr - \frac{c}3 r^2.
\end{equation}
where the explicit form depends on the specific choice of the scale-dependent couplings, and $c$ plays the same role as a cosmological constant. It can be interpreted as the quantum vacuum energy. In the classical limit where $\alpha G$ is constant, the lapse function reduces to the classical solution~\eqref{eq_solution}. However there would be contribution from the quantum effects in general in the intermediate region, so the intermediate behavior may get some distortion.
\end{enumerate}

\subsection{Quantum Improvement at Action Level} \label{sec_action} %\label{subsec:QIAction}
%%%%%%%%%%%%%%%%%%%%%%%%%%%%%%%%%%%%%%%%%%%%%%%%%%%%%%%%%%%%%%%%%%%%%%

An alternative approach is to incorporate quantum effects at the level of the action by promoting the couplings in the classical action~\eqref{eq_action} to the scale-dependent quantities, namely
\begin{equation}
S = \int d^4x \sqrt{- g} \left( \frac{R}{16 \pi G} - \frac{1}{4 e^2} F_{\alpha\beta} F^{\alpha\beta} \right).
\end{equation}
Varying this action with respect to the metric and gauge field yields the Einstein and Maxwell equations
\begin{equation}
\label{eq_action_Ein}
R_{\mu\nu} - \frac12 R g_{\mu\nu} = 8 \alpha G T_{\mu\nu} + \Delta t_{\mu\nu}, \qquad \nabla_\mu \left( \alpha F^{\mu\nu} \right) = 0, 
\end{equation}
where $T_{\mu\nu}$ is the energy-momentum tensor of the Maxwell field defined in~\eqref{eq_MaxwellT}, and $\Delta t_{\mu\nu}$ denotes the effective contribution induced by the scale-dependent Newton coupling:
\begin{eqnarray}
\Delta t_{\mu\nu} &:=& G \left( \nabla_\mu \nabla_\nu - g_{\mu\nu} \nabla_\alpha \nabla^\alpha \right) G^{-1}
\nonumber\\
&=& \frac2{G^2} \nabla_\mu G \nabla_\nu G - \frac1{G} \nabla_\mu \nabla_\nu G - \frac2{G^2} g_{\mu\nu} \nabla_\alpha G \nabla^\alpha G + \frac1{G} g_{\mu\nu} \nabla_\alpha \nabla^\alpha G.
\end{eqnarray}
Note that there is no contribution from the Maxwell field to $\Delta t_{\mu\nu}$ since its action does not have any derivatives of the metric.
Imposing the Bianchi identity, $\nabla_\mu G^{\mu\nu} = 0$, leads to the consistency condition
\begin{equation}
\label{cwb}
8 \alpha G \nabla_\mu T^{\mu\nu} + 8 T^{\mu\nu} \nabla_\mu ( \alpha G ) + \nabla_\mu \Delta t^{\mu\nu} = 0.
\end{equation}
While the first term vanishes due to the conservation of the Maxwell energy-momentum tensor, the second and third terms do not vanish in general. We will argue that it is necessary to introduce a quantum energy-momentum tensor to keep the validity of the Bianchi identity.

On the other hand, to maintain mathematical consistency in the thermodynamics, one has to (i) either modify the action itself, as will be discussed below, or (ii) restrict the functional forms of $G$ and $\alpha$ such that~\eqref{cwb} is satisfied. In the former case, one modifies the action in a covariant manner so that $G$ is treated as a scalar function. This does not break the diffeomorphism invariance and allows one to compute the Noether charge to evaluate the entropy of a stationary black hole solution (if it exists) and derive the first law of thermodynamics by applying the standard covariant phase space method~\cite{Wald:1993nt,Iyer:1994ys}. In the latter case, if $G$ is treated as a fixed background, it breaks the diffeomorphism invariance. Even so, if a stationary black hole solution exists, one can, in principle, examine its thermodynamic consistency by exploiting the stationarity of the background. However, the analysis would be on a case-by-case basis. For example, in the static spherically symmetric case, by restricting $G$ (and $\alpha$) as a function of the area function, one can construct a quantum improved black hole solution which is consistent with the thermodynamic laws as we have discussed in the previous section. For the rotating black hole case, see \cite{Chen:2023wdg}.

We have seen in the previous Subsec.~\ref{sec_equation} that the issue on the Bianchi identity also arises when quantum improvement is implemented directly at the level of the equations of motion.
Now we examine this problem and the compatibility between the quantum improvement at the levels of field equation and action.
To restore the consistency, an additional quantum energy-momentum tensor must be introduced into the Einstein equations such that\footnote{\label{f1} A simple choice is 
$$
\vartheta^\mathrm{(a)}_{\mu\nu} = \vartheta^\mathrm{(e)}_{\mu\nu} - \Delta t_{\mu\nu},
$$
for which the Einstein equations coincide exactly with those obtained at the equation level. However, the Maxwell equations are modified due to the presence of the scale-dependent coupling $\alpha$. }
\begin{equation}
\nabla_\mu \vartheta^\mathrm{(a)}{}^{\mu\nu} = - 8 T^{\mu\nu} \nabla_\mu ( \alpha G ) - \nabla_\mu \Delta t^{\mu\nu}.
\end{equation}
The trace and covariant divergence of the tensor $\Delta t_{\mu\nu}$ are given by
\begin{equation}
\Delta t := \Delta t^\mu{}_\mu = \frac3{G} \nabla_\alpha \nabla^\alpha G - \frac6{G^2} \nabla_\alpha G \nabla^\alpha G, \qquad \nabla_\mu \Delta t^{\mu\nu} = \frac1{G} \nabla_\mu G \left( \Delta t^{\mu\nu} - R^{\mu\nu} \right).
\end{equation}
Using the quantum-improved Einstein equations, with the quantum energy–momentum tensor $\vartheta^{\mathrm{(a)}}_{\mu\nu}$ incorporated into equation~\eqref{eq_action_Ein}:
\begin{equation}
\label{eq_action_QEin}
R_{\mu\nu} = 8 \alpha G \left( T_{\mu\nu} - \frac12 T \, g_{\mu\nu} \right) + \Delta t_{\mu\nu} - \frac12 \Delta t \, g_{\mu\nu} + \vartheta^{\mathrm{(a)}}_{\mu\nu} - \frac12 \vartheta^{\mathrm{(a)}} \, g_{\mu\nu},
\end{equation} 
where $T := T^\mu{}_\mu$ and $\vartheta^{\mathrm{(a)}} := \vartheta^{\mathrm{(a)}\mu}{}_\mu$, we obtain the on-shell consistency condition
\begin{eqnarray}
\nabla_\mu \vartheta^{\mathrm{(a)}\mu\nu} - \frac{\nabla_\mu G}{G} \left( \vartheta^{\mathrm{(a)}\mu\nu} - \frac12 \vartheta^{\mathrm{(a)}} \, g^{\mu\nu} \right) &=& - 8 T^{\mu\nu} \nabla_\mu (\alpha G) + 8 \alpha \nabla_\mu G \left( T^{\mu\nu} - \frac12 T g^{\mu\nu} \right)
\nonumber\\
&& - \nabla^\nu G \left( \frac3{2 G^2} \nabla_\alpha \nabla^\alpha G - \frac3{G^3} \nabla_\alpha G \nabla^\alpha G \right).
\label{cons}
\end{eqnarray}
In the vacuum case $T_{\mu\nu} = 0$, it was argued in~\cite{Reuter:2003ca} that the exact solution is uniquely fixed to the form corresponding to Brans–Dicke theory
\begin{equation}
\vartheta^{\mathrm{(a)}}_{\mu\nu}\Big|_{T_{\mu\nu}=0} = \vartheta^{\mathrm{BD}}_{\mu\nu} := - \frac3{2 G^2} \left( \nabla_\mu G \nabla_\nu G - \frac12 g_{\mu\nu} \nabla_\alpha G \nabla^\alpha G \right).
\end{equation}
It is noted that this solution is unique only if one demands that~\eqref{cons} is valid for {\it all} backgrounds $G(x)$~\cite{Reuter:2003ca}.
The solution remains valid in the presence of a traceless source ($T = 0$) provided that the electromagnetic coupling does not run, i.e., $\alpha = \pi/e_0^2$ is constant. In this case, the quantum-improved action takes the form
\begin{equation} 
S = \frac1{16 \pi} \int d^4x \sqrt{-g} \left( \frac{R}{G} + \frac3{2 G^3} \nabla_\alpha G \nabla^\alpha G - \frac{4\pi}{e_0^2} F_{\alpha\beta} F^{\alpha\beta} \right).
\end{equation}
Taking the trace of the quantum-improved Einstein equation~\eqref{eq_action_QEin}, we obtain
\begin{equation}
R = - \frac3{G} \nabla_\alpha \nabla^\alpha G + \frac9{2 G^2} \nabla_\alpha G \nabla^\alpha G.
\end{equation}
One can readily verify that the equation obtained from variation with respect to $G$,
\begin{equation}
\nabla_\alpha \nabla^\alpha G - \frac3{2 G} \nabla_\alpha G \nabla^\alpha G + \frac13 G R = 0,
\end{equation}
does not give an independent equation of motion, but instead reproduces the trace of the Einstein equations. Consequently, the scale-dependent coupling $G$ remains undetermined.

This is to be expected since $G$ is originally just a coupling which gets scale dependence from the quantum effects but not dynamical field.
This fact may be most easily confirmed if we go to the Einstein frame. Setting the metric to
\bea
g_{\mu\nu} = \frac{G(x)}{G_0} \tilde g_{\mu\nu},
\eea
we get the action in the Einstein frame:
\bea
S_E = \frac1{16 \pi} \int d^4x \sqrt{-\tilde g} \left( \frac{\tilde R}{G_0} - \frac{4\pi}{e_0^2} F_{\alpha\beta} F^{\alpha\beta} \right),
\eea
where it should be understood that the contraction is made with respect to the metric $\tilde g_{\mu\nu}$. Thus we see that if we add the additional ``quantum term'' to the action, the scale-dependent coupling completely disappear and does not have any dynamics in the Einstein frame. Since physics does not change by the choice of frames, we conclude that $G$ is not a dynamical field. Unfortunately this means that there is no scale-dependent Newton coupling $G$ in the Einstein frame and no quantum effects.

We emphasize that the above result is obtained under the assumption that~\eqref{cons} is satisfied for {\it all} backgrounds $G(x)$. If we relax this condition, it is possible that the result in the action level becomes consistent with the approach of quantum improvement at the equation level in Subsec.~\ref{sec_equation}. Indeed if we can choose
\begin{equation} 
\vartheta^\mathrm{(a)}_{\mu\nu} = \vartheta^\mathrm{(e)}_{\mu\nu} - \Delta t_{\mu\nu},
\end{equation}
the Einstein equations coincide exactly with those~\eqref{eq_vartheta_e} obtained at the equation level (See Eq.~\eqref{eq_action_QEin} and footnote~\ref{f1}).
%\bea
%R_{\mu\nu} - \frac12 R g_{\mu\nu} = 8 \alpha G T_{\mu\nu} + \vartheta^\mathrm{(e)}_{\mu\nu}.
%\eea
The Einstein equations would follow from the action of the form
\bea
S = \frac1{16 \pi} \int d^4x \sqrt{-g} \left[ \frac{R}{G} + \frac{1}{G^3} \left( 2 \nabla_\alpha G \nabla^\alpha G - G \nabla_\alpha \nabla^\alpha G \right) - \frac{4 \pi}{e^2} F_{\alpha\beta} F^{\alpha\beta} \right] + S_{\vartheta}[g,G],
\label{modaction}
\eea
where the last term is the action giving $\vartheta^\mathrm{(e)}_{\mu\nu}$. We expect that such term arises from the quantum effects so as to make the system consistent.
For such an additional quantum energy-momentum tensor, we may have the action
\bea
S_{\vartheta}[g, G] = \int d^4x \sqrt{-g} \, \frac1{16 \pi G} g^{\mu\nu} \vartheta^\mathrm{(e)}_{\mu\nu},
\eea
which gives the desired term.
A condition for the validity of $S_\vartheta[g, G]$ is that the solution for $\vartheta^\mathrm{(e)}_{\mu\nu}$ does not explicitly depend on the metric.
This is the case, for example, if it is the energy-momentum tensor for the perfect fluid.
If these results are valid, the modifications due to quantum effects can make this approach equivalent to the quantum improvement considered in Subsec.~\ref{sec_equation}.

Actually the result is almost correct but not quite. We find that there is a difference in the trace part of the terms originating from the term depending on $G$. One possible resolution of the problem is, for example, that the trace terms cancel out:
\bea
2 \nabla_\alpha G \nabla^\alpha G = G \nabla_\alpha \nabla^\alpha G,
\eea
which is equivalent to the massless scalar field equation:
\bea
\nabla_\alpha \nabla^\alpha \left(\frac{1}{G}\right)=0.
\label{solution1}
\eea
When we are concerned with the behavior of $G$ as a function of the area radius on and outside the event horizon, the operator $\nabla_\alpha \nabla^\alpha$ would become elliptic, and $1/G$ should be a harmonic function.
It may appear then that the additional action in~\eqref{modaction} constructed from derivatives of $G$ vanish if $1/G$ is a harmonic function. However, the variation should be performed prior to imposing the consistency condition on $G$. In this case, $\Delta t_{\mu\nu}$ is traceless but the action does not vanish.

For the spherically symmetric spacetime~\eqref{eq_metric} with a point charge source~\eqref{qMaxwell}, let us assume that the quantum energy–momentum tensor takes the perfect-fluid form~\eqref{perf_f}.
We get, from Eq.~\eqref{solution1}, together with the consistency condition~\eqref{eq_vartheta_e} and the Einstein equation~\eqref{eq_QfieldEQ}, a system of coupled equations for the Newton coupling $G(r)$, the quantum energy density $\rho_q(r)$, and the metric function $f(r)$:
\begin{eqnarray}
r^2 f G' - c_1 G^2 &=& 0,
\label{eq1}
\\
r^4 \rho_q' + 2 (1 + w) r^3 \rho_q + Q^2 (\alpha G)' &=& 0,
\label{eq2}
\\
r^3 f' + r^2 f - r^2 + r^4 \rho_q + Q^2 \alpha G &=& 0.
\label{eq3}
\end{eqnarray}
with an integration constant $c_1$.
This is the {\it complete} set of equations that give solutions for $G(r), \rho_q(r)$ and $f(r)$ in our solution at the level of action. 
%, consistent with the quantum improvement at the level of equations.
Note that Eq.~\eqref{eq2} is the same as~\eqref{eq02} in the quantum improvement at the equation level, and~\eqref{eq3} is also the same Einstein equation. The only difference is~\eqref{eq1}, which is absent at the equation level. Thus the quantum improvement at the action level is equivalent to that at the equation level on the assumption~\eqref{eq1}.

In the simplest case $Q = 0$, we can find the general solution: 
\begin{equation}
\rho_q = c_2 r^{-2 w - 2}, \qquad f = 1 - \frac{c_3}{r} + \frac{c_2}{2 w - 1} r^{-2 w}, 
\end{equation} 
and 
\begin{equation}
\frac1{G} = c_4 - \int \frac{c_1}{r^2 f} dr,
\end{equation}
where $c_2, c_3$ and $c_4$ are again integration constants.
If we write $c_3 = 2 G_0 M$ and $c_4 = 1/G_0$, $G_0$ is the low-energy value of the Newton coupling and $M$ is the mass of the black hole.
The resulting geometry corresponds to the Kiselev black hole~\cite{Kiselev:2002dx}. In the present context, the source of this solution comes from quantum effects.

Let us consider some simple situations.

\begin{enumerate}
\item [(i)]
If $c_2 = c_3 = 0$, the space is flat Minkowski space with $f = 1$ and $\rho_q = 0$. This gives
\bea
\frac{1}{G} = \frac{1}{G_0} - \frac{c_2}{r}.
\eea
In the asymptotic region $r \to \infty$, $G$ goes to the constant $G_0$. This is a reasonable behavior since the quantum effects becomes small there. In the short distance limit $r \to 0$, $G$ vanishes due to the quantum effects.

\item [(ii)]
If we consider the standard Schwarzschild solution like $f = 1 - c_3/r$ with $c_2 = 0$, Eq.~\eqref{eq1} gives
\bea
\frac{1}{G} = \frac{1}{G_0} + \frac{c_1}{c_3} \log\left|\frac{r-c_3}{r}\right|.
\eea
Again in the asymptotic region, $G$ goes to the constant $G_0$. When we come close to the ``horizon'' at $r=c_3$, the logarithm diverges and the Newton coupling vanishes. It remains to be seen if this is a reasonable behavior. After that, in the short distance limit, it again becomes zero.

%\item [(iii)]
%Actually in the above case (ii), $c$ may depend on $G$ and this gives nonlinear equation to be solved. We have solve our complete set of Eqs.~\eqref{eq1} -- \eqref{eq3}.
\end{enumerate}

For $Q \neq 0$, general solutions for $\rho_q$ and $f(r)$ are already given in Eqs.~\eqref{sol1} and~\eqref{sol2} in the preceding subsection. In these solutions, $G(r)$ is involved on the rhs of the equations, and it has to be determined by Eq.~\eqref{eq1}, which contains $f(r)$. Thus this is a complicated system of coupled equations, in contrast to the quantum improvement at the equation level. It is, however, possible to numerically integrate the differential equations to find black hole solutions for given $Q^2, c_1$ and $w$ once suitable boundary conditions are fixed.

There may be also other possibilities with other energy-momentum tensor. It would be interesting to study if there is any other reasonable solutions.

%%%%%%%%%%%%%%%%%%%%%%%%%%%%%%%%%%%%%%%%%%%%%%%%%%%%%%%%%%%%%%%%%%%%%%
\section{Bianchi-type-I Cosmological solutions with radiation} \label{sec_cosmology}
%%%%%%%%%%%%%%%%%%%%%%%%%%%%%%%%%%%%%%%%%%%%%%%%%%%%%%%%%%%%%%%%%%%%%%

We note that the interior of the static black hole~\eqref{eq_metric}, i.e., the region where $f(r) < 0$, may be regarded as describing the Kantowski-Sachs type universe~\cite{Kantowski:1966te}. Furthermore, if we replace the unit sphere part of~\eqref{eq_metric} with the flat geometry, it can describe the Bianchi-type-I universe. Let us focus on the region $f(r) < 0$, rewrite $r \rightarrow a$, $-f \rightarrow b^2$, $t \rightarrow x$, and $d\Omega^2 \rightarrow dy^2 + dz^2$, in~\eqref{eq_metric}, and then we obtain the Bianchi-I metric
\begin{eqnarray}
ds^2 = - \dfrac{da^2}{b^2} + b^2(\tau) dx^2+ a^2(\tau) (dy^2 + dz^2)  \,.     
\end{eqnarray}
The cosmic proper time $\tau$ is $\tau = \int da/b$, but to keep our discussion as analogous to the black hole case as possible, we identify $a$ with our time coordinate.  
If we consider the uniform electric field along $x$-axis 
\begin{eqnarray}
A = \dfrac{Q}{a} dx \,, \qquad F_{ax}= - \dfrac{Q}{a^2} \,, 
\end{eqnarray}
with constant $Q$, the energy-momentum tensor is
\begin{eqnarray}
T^\mu{}_\nu = \dfrac{Q^2}{2 a^4} {\rm diag} (-1,-1,1,1) \,.
\end{eqnarray}
On the other hand, the uniform magnetic field along $x$-axis
\begin{eqnarray}
A = B (-y dz + z dy) \,, \qquad F_{yz} = - B \,, 
\end{eqnarray}
with $B$ constant, gives the energy-momentum tensor again the same form
\begin{eqnarray}
T^\mu{}_\nu = \dfrac{B^2}{2 a^4} {\rm diag} (-1,-1,1,1) \,.
\end{eqnarray}
This is just a manifestation of the duality in the system~\cite{Chen:2026ntp}.

\subsection{Quantum improvement at equation level}
%%%%%%%%%%%%%%%%%%%%%%%%%%%%%%%%%%%%%%%%%%%%%%%%%%%%%%%%%%%%%%%%%%%%%%

The scale-dependent couplings $G, \alpha$ appear in the same manner in the present case. We need to introduce 
\begin{eqnarray}
\vartheta^\mathrm{(e)}{}^\mu{}_{\nu} = {\rm diag} \Bigl( -\rho(a),-\rho(a), w \rho(a), w \rho(a)  \Bigr),
\end{eqnarray}
with
\begin{eqnarray}
\rho(a) = a^{-2(w+1)} \left( c - \int da \dfrac{Q^2}{a^{-2(w-1)}} (\alpha G)' \right).  
\end{eqnarray}
Just like~\eqref{sol2}, we have
\begin{eqnarray}
b^2 = \dfrac{C}{a} +\dfrac{1}{a} \int \left( \dfrac{Q^2 \alpha G}{a^2} + a^2 \rho(a) \right) da.
\end{eqnarray}
Note that constant $C$ corresponds to $2 G_0 M$, and the first term $+1$ in~\eqref{sol2} disappears as we have replaced the unit-sphere metric $d\Omega^2$ with the flat metric.  
If we further assume that $\vartheta^\mathrm{(e)}_{\mu \nu}$ takes the cosmological-constant-type as~\eqref{theta:cc-type}, we obtain 
\begin{eqnarray}
b^2 = + \dfrac{C}{a} - \dfrac{4 Q^2}{a} \int a^2 \int^a \dfrac{\alpha(a')G(a')}{a'^5} da' da + \dfrac{c}{3} a^2.
\end{eqnarray}
From this expression, we find that when $a$ is small, $b$ can exhibit behavior significantly different from that of $a$. However, once $a$ becomes sufficiently large, the last term begins to dominate, causing $b$ to grow proportionally to $a$. As a result, the Bianchi universe approaches an isotropically expanding universe. Therefore, the above model qualitatively reproduces the same isotropization behavior induced by quantum gravity effects as discussed in Refs.~\cite{Chen:2025ybu, Chen:2026ntp}.

\subsection{Quantum improvement at action level}
%%%%%%%%%%%%%%%%%%%%%%%%%%%%%%%%%%%%%%%%%%%%%%%%%%%%%%%%%%%%%%%%%%%%%%

The basic arguments follow from those of Subsec.~\ref{sec_action}. If we assume that $G$ obeys~\eqref{solution1}, and the quantum stress-energy tensor does the form~\eqref{perf_f}, by replacing 
$r \rightarrow a$, $-f \rightarrow b^2$, $t \rightarrow x$, and $d\Omega^2 \rightarrow dy^2 + dz^2$ in~\eqref{eq1},~\eqref{eq2} and~\eqref{eq3}, we obtain
\begin{eqnarray}
a^2 b^2 \dfrac{dG}{da} + c_1 G^2 &=& 0,
\label{eq:G}
\\
a^4 \dfrac{d \rho_q}{da} + 2 (1 + w) a^3 \rho_q + Q^2 \dfrac{d(\alpha G)}{da} &=& 0,
\\
a^3 \dfrac{d(b^2)}{da} + a^2 b^2 - a^4 \rho_q - Q^2 \alpha G &=& 0.
\label{eq:b2}
\end{eqnarray}
The third term in~\eqref{eq3} is removed in~\eqref{eq:b2} as we replaced the unit positive curvature space $d\Omega^2$ with the vanishing curvature space $dy^2+dz^2$.

If we further assume, for simplicity, $Q = 0$, we have (for the case $w \neq 1/2$), 
\begin{equation}
\rho_q = c_2 a^{-2(1+w)} \,, \qquad
b^2 = \dfrac{c_3}{a} + \dfrac{c_2}{1-2w} a^{-2w}\,, \qquad
\dfrac{1}{G} = c_4 + c_1 \int\dfrac{da}{a^2 b^2}  \,.
\label{eq:b2s0} 
\end{equation}
Note that, as the constant $c_3$ corresponds to the mass term in the black hole case~\eqref{sol2}, it measures the magnitude of the Weyl curvature.  
In particular, we find 
\begin{eqnarray}
\dfrac{1}{G} 
= c_4 + \dfrac{c_1}{c_3 (1 - 2 w)} \log\left| \dfrac{a^{1-2w}}{c_2 a^{1-2w} + c_3 (1-2w)} \right|.
\end{eqnarray}
It follows that provided $c_1,c_2,c_3$ and $c_4$ are chosen appropriately, $G$ asymptotically ($a \rightarrow \infty$) approaches the constant $G_0$ if $w<1/2$, whereas $G$ asymptotically vanishes if $w>1/2$.
In order for $a$ to serve as a time coordinate, $b^2$ must remain nonnegative. We then see from Eq.~\eqref{eq:b2s0} that $c_2$ and $c_3$ should be positive and $w < 1/2$.

When $w=1/2$, we have to go back to Eqs.~\eqref{eq:G}--\eqref{eq:b2} to find that $\rho_q =c_2a^{-3}$ and $b^2$ is given by 
\begin{eqnarray}
b^2 = \dfrac{c_3}{a} + c_2 \dfrac{\log a}{a} \,.
\label{eq:b2s} 
\end{eqnarray}
Then, we obtain
\begin{eqnarray}
\dfrac{1}{G} = c_4 + \dfrac{c_1}{c_2} \log\left| c_2 \log a + c_3 \right|.
\label{eq:G2}
\end{eqnarray}
In this case, although $G$ continues to evolve toward a vanishingly small value with the expansion of the universe, its rate of change is extremely small.
Again $b^2$ should be positive, and we see from Eq.~\eqref{eq:b2s} that $c_2$ and $c_3$ should be positive.
We also require that the Newton coupling $G>0$, which implies that $c_1>0$ and $c_4>0$ according to Eq.~\eqref{eq:G2}.

It is intriguing to point out the following possibility. Suppose that the Weyl curvature $c_3$ is non-vanishing, 
%$c_2$ is a positive constant, 
and the quantum matter behaves as dust ($w=0$). In this setup, we can consider the $yz$-plane as a 2-dimensional expanding universe with scale factor $a$, and the $x$-direction as an extra-dimension. As the 2-dimensional universe expands from the big bang to asymptotic future ($a\rightarrow \infty$), the scale factor of the extra-dimension, $b$, asymptotes from infinity to the constant value $c_2$. Simultaneously, $G$, which is initially very small, approaches a constant value. 
Such a behavior of the two scale factors, $a$ and $b$, implies that 
the $2$-dimensional homogeneous and isotropic space is expanding, while 
1-dimensional extra-dimension with $b$ approaches a constant and stabilized as the universe expands. 
Now let us generalize this idea to a $5$-dimensional spacetime which consists of a $4$-dimensional expanding homogeneous and isotropic universe with scale factor $a$ and one extra-dimension compactified on a circle with scale factor $b$. 
One can expect the similar behavior of $a$ and $b$ mentioned above.
This observation is expected to remain valid even for exponential growth of $a$. Such a phenomenon is known as the Kaluza-Klein inflation, demonstrated first by the higher dimensional cosmological model with a dust fluid~\cite{Ishihara:1984wx}. However, the present case differs from the known Kaluza-Klein inflation models in the following two key respects. First, the dust-like fluid here originates from quantum effects, and second, the value of $G$ itself vanishes near the big bang.

%For example, when the dust type quantum matter case ($w = 0$), 
%\begin{equation}
%\dfrac{1}{G} = c_4 + \dfrac{c_1}{c_3} \log\left( \dfrac{a}{c_2 a + c_3} \right). 
%\end{equation}
%When the cosmological constant type case ($w = -1$), we have
%\begin{equation}
%\dfrac{1}{G} = c_4 + \dfrac{c_1}{3 c_3} \log\left( %\dfrac{3a^3}{c_2 a^3 + 3 c_3} \right). 
%\end{equation}

%%%%%%%%%%%%%%%%%%%%%%%%%%%%%%%%%%%%%%%%%%%%%%%%%%%%%%%%%%%%%%%%%%%%%%
\section{Conclusion}
\label{sec_conclusion}
%%%%%%%%%%%%%%%%%%%%%%%%%%%%%%%%%%%%%%%%%%%%%%%%%%%%%%%%%%%%%%%%%%%%%%

We have investigated the quantum-improved charged black holes within the framework of asymptotically safe gravity, with particular emphasis on the consistency of black hole thermodynamics, the compatibility of the quantum-improved field equations with the Bianchi identity and the consistency between the equation and action levels.

The first problem is concerned with the thermodynamics.
Unlike many previous studies on regular rotating and/or charged black holes, where thermodynamic consistency is often assumed implicitly, we have shown that it provides a strong and nontrivial constraint on viable quantum-improved solutions.

At the solution level, we have considered the RN black hole with both the Newton and electromagnetic couplings promoted to the scale-dependent quantities. We have demonstrated that thermodynamic consistency is preserved provided that the scale-dependent couplings depend only on the horizon radius, or equivalently on the horizon area. In particular, the charge and the scale-dependent electromagnetic coupling appear in the combinations $Q^2/e(r)^2$ and $Q/e(r)^2$, respectively, which guarantees the integrability of the first law. This result also implies that regular charged black holes constructed by introducing an arbitrary radial charge distribution generally fail to preserve thermodynamic consistency. Our analysis therefore provides an important criterion for evaluating the physical viability of regular charged black hole solutions.

The second problem is the compatibility of the quantum-improved field equations with the Bianchi identity.
We have investigated quantum improvement at the level of the field equations. Since the scale-dependent couplings spoil the standard conservation properties implied by the Bianchi identity, it becomes necessary to introduce an additional quantum energy-momentum tensor. Under suitable assumptions on its structure, we derived consistent quantum-improved Einstein equations and obtained explicit forms of the corresponding black hole solutions. Depending on the equation of state of the quantum energy-momentum tensor, the resulting geometry can reproduce either RN-type behavior or effective cosmological-constant-like contributions generated by quantum effects.

The third problem is the consistency between the approaches at the equation and action levels. At the action level, we have examined the consistency conditions with the equation level arising from promoting the couplings directly in the Einstein-Maxwell action. It had been argued that if one requires consistency for arbitrary background couplings, the resulting theory reduces effectively to a Brans-Dicke type theory~\cite{Reuter:2003ca} whose Einstein-frame description eliminates the scale-dependent Newton coupling and hence removes the quantum effects. However, by relaxing this overly restrictive requirement and allowing consistency only for a given scale-dependent coupling, we find that it becomes possible to construct a consistent quantum-improved framework that is consistent with the equation-level improvement. In this case too, the additional quantum energy-momentum tensor plays an essential role in restoring consistency.

We also studied the cosmological interpretation of the interior region of the quantum-improved charged black holes. By rewriting the interior geometry as a Bianchi-I cosmology, we found that the quantum corrections can drive the universe toward isotropic expansion at late times. In particular, for cosmological-constant-type quantum energy-momentum tensors, the anisotropic scale factors asymptotically become proportional to each other, qualitatively reproducing the isotropization behavior induced by quantum gravity effects found in previous works.

Several important issues remain to be explored. In particular, a more detailed analysis of thermodynamic consistency for regular charged black holes constructed within nonlinear electrodynamics would be highly desirable. 
Another important issue is what is the general solution to the consistency of the approaches between the equation and action levels.
We hope that these consistency conditions identified in this work will provide useful guidance toward the construction of physically viable quantum-improved black hole solutions.

\acknowledgments

C.M.C. would like to thank Kimet Jusufi for valuable discussions.
The work of C.M.C. was supported by the National Science Council of the R.O.C. (Taiwan) under the grants NSTC 114-2112-M-008-010 and 115-2918-I-008-006.
N.O. would like to thank Department of Physics, Nagoya University for the hospitality during his visit, where part of this work is done.
The work of A.I. was in part supported by JSPS KAKENHI Grant No. JP25K07306, JP26K07105 and also supported by MEXT KAKENHI Grant-in-Aid for Trans-
formative Research Areas A Extreme Universe No. JP21H05182 and JP21H05186.

\appendix

\section{Comparison between the quantum improvements at the solution and equation levels}

In this appendix, we compare the two procedures for quantum improvements at the solution and equation levels. In~\cite{Ishibashi:2021kmf}, the quantum improvement of a static charged black hole was considered at the solution level. Strictly speaking, the coupling constant of the electromagnetic field runs through its interaction with matter fields. Since our current setup does not incorporate such matter couplings, it differs from that of the previous work~\cite{Ishibashi:2021kmf}. For this reason, one should be cautious about drawing immediate conclusions from a simple comparison. Nevertheless, we believe that by such a comparison, one may highlight the differences between the two methods.  

As given in~\cite{Ishibashi:2021kmf}, the two running couplings with energy scale $k$ are found to behave as 
\begin{equation}
G(k) = \dfrac{G_0}{1 + G_0k^2/(4 \pi \tilde{\alpha})} \,, 
\end{equation}
and 
\begin{eqnarray}
1/e^2(k) &=& C_0 (1 + D k^2)^{\tilde\alpha} + \dfrac{1}{e_*^2} \dfrac{\tilde\alpha}{(1 - \tilde\alpha)} (1 + D k^2) F(1, 1 - \tilde\alpha, 2 - \tilde\alpha; 1 + D k^2) \,,
\qquad \mbox{for} \quad {\tilde \alpha \neq 1}
\\
1/e^2(k) &=& \left( 1 + \dfrac{G_0}{4\pi} k^2 \right) \left\{ C_0 - \dfrac{1}{e_*^2} \log\left( \dfrac{G_0k^2}{4 \pi + G_0 k^2}\right) \right\},
\qquad \mbox{for} \quad {\tilde \alpha = 1} 
\end{eqnarray}
where $F$ denotes the hypergeometric functions and $C_0$ is a constant and $D = G_0/4\pi {\tilde \alpha}$. Here the parameter $\tilde \alpha$ specifies the UV fixed points of the dimensionless coupling constants $G_*, e_*$ (with the additional parameter $b$ (see Eqs. (3.2)-(3.4) in Ref.~\cite{Ishibashi:2021kmf}) and is assumed to be $0<{\tilde \alpha}\leq 1$. Let us restrict our attention to the ${\tilde \alpha} \neq 1$ case. The scale identification was made by using the Kretschmann scalar $K(r)$ of the classical solution of the Reissner-Nordstr\"om metric so that $k^2 \propto K(r)$. Then, the couplings behave as, in UV $k^2 \sim 1/r^4$, 
\begin{equation}
G \simeq A {\tilde \alpha} r^4 + O(r^5) \,, \qquad 
e^2 \simeq 4 \pi B {\tilde \alpha} r^{4 {\tilde \alpha}} + O(r^{4{\tilde \alpha}+1}) \,, 
\end{equation}
with $A, B$ some positive constants, while in IR $k^2 \sim 1/r^3$,  
\begin{equation}
G \simeq G_0 \,, \qquad e^2 \simeq \dfrac{e_*^2}{{\tilde \alpha} \log(1/Dk^2)} \simeq \dfrac{e_*^2}{{\tilde \alpha} \log(r^3/D)} \,.
\end{equation}

\begin{enumerate}
\item{\bf Quantum improvement at solution level:} 

The metric function $f(r)=-g_{tt}$ behaves as follows:
\begin{itemize}
\item
In UV limit ($r\rightarrow 0)$, the asymptotic behavior of the metric function toward the center becomes
\begin{equation}
f(r) \simeq 1 -2MA r^3 + AB r^{4{\tilde \alpha}+2} \,, 
\end{equation}
implying the resolution of the central singularity.

\item
In IR limit ($r \rightarrow \infty$), the asymptotic behavior of the metric function at large distances becomes 
\begin{equation}
 f(r) \simeq 1-\dfrac{2G_0M}{r}+ C_s\cdot \dfrac{G_0 Q^2}{r^2} \cdot \dfrac{1}{\log r}  
\end{equation}
with $C_s$ some positive constant. Although the metric approaches asymptotically flat, due to the logarithmic dependence of the last term, the Coulomb potential term decays slightly faster than the classical Reissner-Nordstrom case. 
\end{itemize}

\item{\bf Quantum improvement at equation level:}  

Now let us apply the same Kretschmann scale identification to the quantum improvement at equation level. 

\begin{itemize}
\item{In UV limit ($r\rightarrow 0)$} (i.e., $k^2 \sim 1/r^4$), substituting the above behavior of $G, e^2$ into~\eqref{lapse_QI_eqlevel}, we find:  
\begin{equation}
 f(r) \simeq 1- \dfrac{2G_0M}{r} + C_u\cdot Q^2 \cdot r^{4{\tilde \alpha}+2} \,, 
\end{equation} 
where $C_u$ is some positive constant. Thus the qualitative behavior of the Coulomb term is the same as the solution level improvement. 
The difference from the solution level is manifest in the second gravitational potential term, which produces the singularity at the center.  

\item{In IR limit ($r \rightarrow \infty$)} (i.e., $k^2 \sim 1/r^3$), Substituting $G \rightarrow G_0$ and $e^2 \propto e_*^2/\log(r)$ into~\eqref{lapse_QI_eqlevel}, we find that at large distances, 
\begin{equation}
f(r) \simeq 1- \dfrac{2G_0M}{r} + C_e\cdot \dfrac{G_0 Q^2}{r^2} \cdot \log r\,, 
\end{equation} 
with $C_e$ some positive constant. Thus, although the metric asymptotically approaches the flat metric, the Coulomb term decays slower than its classical counterpart and also than that of the solution level case. 

\end{itemize} 

\end{enumerate}

\end{document}